\documentclass[12pt,prd,aps,onecolumn,nofootinbib]{revtex4-1}%
\usepackage[colorlinks=true,linkcolor=blue,urlcolor=blue,filecolor=black,citecolor=red,pdfstartview=FitV,pdftitle={},pdfsubject={},
pdfkeywords={},pdfpagemode=None,bookmarksopen=true]{hyperref}
\usepackage{graphicx}
\usepackage{epstopdf}%
\usepackage{amsmath}
\usepackage{amsfonts}
\usepackage{amssymb}
\usepackage{xcolor}
\usepackage{color}%
\usepackage{dcolumn}
\usepackage{slashed}
\usepackage[normalem]{ulem}
\usepackage{amsthm}
\usepackage{mathrsfs}
\usepackage{subfigure}
\usepackage{wrapfig}
\usepackage{indentfirst}

\providecommand{\U}[1]{\protect\rule{.1in}{.1in}}

\begin{document}

\title{Gravitational lensing effect of black holes in effective quantum gravity}

\author{Hao Liu$^{1}$}
\author{Meng-Yun Lai$^{1}$}~\email[]{mengyunlai@jxnu.edu.cn (corresponding author)}
\author{Xiao-Yin Pan$^{2}$}~\email[]{panxiaoyin@nbu.edu.cn}
\author{Hyat Huang$^{1}$}~\email[]{hyat@mail.bnu.edu.cn}
\author{De-Cheng Zou$^{1}$}~\email[]{dczou@jxnu.edu.cn}

\affiliation{$^{1}$College of Physics and Communication Electronics, Jiangxi Normal University, Nanchang 330022, China}
\affiliation{$^{2}$Department of Physics, Ningbo University, Ningbo, Zhejiang 315211, China}

\begin{abstract}
In the present work, we investigate the gravitational lensing effects of two quantum-modified black hole models recently proposed in effective quantum gravity.  {The light deflection angles are calculated for both the weak-field and strong-field limits.} Furthermore,  {using the data for the supermassive black holes SgrA* and M87*, we calculate the lensing observables in the strong-field limit}.  {We find that the quantum parameter plays a role analogous to the electric charge in weak gravitational lensing. In the strong-field limit, in contrast, the effects of the quantum parameter on the deflection angle, the angular separation, and the relative magnification are opposite to those of the electric charge, the scalar charge, and the quantum parameters in some gravity theories.} The results indicate the  {crucial difference} between the classical black holes and the two quantum-modified black hole models that depend on the quantum correction, making them a valuable tool for  {distinguishing} these black hole models.
\end{abstract}

\maketitle








\section{Introduction}\label{1s}

 {General Relativity (GR), as a well-established classical gravitational theory \cite{Will:2014kxa,LIGOScientific:2016aoc}, remains widely regarded as our best description of how spacetime behaves on macroscopic scales to date. However, despite the remarkable successes of GR, it faces significant challenges, such as its inconsistency with quantum physics \cite{Polchinski:1998rq,Ambjorn:2001cv,Ashtekar:2015vfx} and the existence of singularities \cite{Penrose:1964wq,Olmo:2017fbc,hawking2023large}, including the big bang cosmological singularity and black hole (BH) singularities. This suggests that GR might not be the ultimate theory of spacetime and prompts ongoing efforts to modify Einstein's theory. For example, non-local theories with higher derivatives, alternative theories with extra fields, and higher dimensional theories of gravity have been explored. For reviews, see 
\cite{Clifton:2011jh,Mukohyama:2019unx,Doneva:2022ewd} and references therein.}

 {Even though a comprehensive theory and experimental evidence on the quantum aspects of gravity are still lacking \cite{Ashtekar:2021kfp}, it has long been suggested that the singularity problem should be resolved by the quantum theory of gravity (for modern quantum gravity, see \cite{Rovelli:2004tv,Thiemann:2007pyv,Percacci:2023rbo} and references therein). To explore quantum gravity (QG) effects, there are several approaches. For instance, a fundamental length scale is proposed in non-commutative geometry \cite{Nicolini:2005vd}, which smooths out the singularities. Donoghue \cite{Donoghue:1994dn} obtains modified metrics by adding higher-derivative terms to the gravitational action. The AdS/CFT correspondence translates the quantum corrections from conformal field theory to modifications in the AdS BH metric \cite{Maldacena:1998zhr}. Moreover, lots of efforts are devoted to describing some phenomena at energy scales  much lower than the Planck scale and providing possible quantum corrections to GR, so-called  effective field theory (EFT) of gravity \cite{buchbinder2017effective,Donoghue:1994dn,Burgess:2003jk}. Besides, there exists a quite different approach, the Hamiltonian constraints approach \cite{Thiemann:2007pyv,Ashtekar:2004eh,Yao:2023qjd}. In this approach, one abandons the action in the 4-dimensional metric tensor. Consequently, the Einstein equations is reformulated within the Hamiltonian framework and the dynamics are governed by the constraints. In particular, the Hamiltonian approach is important in semiclassical models derived from canonical quantum gravities, such as Loop Quantum Gravity (LQG)
\cite{Ashtekar:2021kfp,Perez:2017cmj,Lewandowski:2022zce,Pan:2022pjh,Han:2024reo}. LQG has been widely applied to cosmology and black hole physics with considerable success \cite{Ashtekar:1997yu,Bojowald:2001xe,Ashtekar:2003hd,Boehmer:2007ket,Chiou:2008nm,Saini:2017ggt,Saini:2018tto,Bodendorfer:2019cyv,Alonso-Bardaji:2021yls,Alonso-Bardaji:2022ear,Yang:2022btw,Zhang:2023okw,Shao:2023qlt}.}


 {Recently, two new black hole models are proposed in Ref. \cite{Zhang:2024khj} by considering general covariance within the Hamiltonian approach (see, e.g. \cite{Bojowald:2011aa,Tibrewala:2013kba,Bojowald:2015zha,Wu:2018mhg,Bojowald:2020unm,Gambini:2022dec,Bojowald:2024beb} for general covariance in canonical models). The work \cite{Zhang:2024khj} has attracted attentions. The quasinormal modes, light rings and shadows of these quantum-modified BH models have been investigated \cite{Konoplya:2024lch,Liu:2024soc}. However, the gravitational lensing effects of these two models still lack for investigation, which will be the aim of this work.}

Gravitational lensing  {ocurrs} when the light is deflected by a massive body or concentration of matter, such as cluster of galaxies,  {inducing} a gravitational field that causes substantial curvature in spacetime, leading to the bending of light and changes in the position and characteristics of the source as observed from Earth. The deflection of light is one of the most important predictions of GR, and played a key role as a first experimental verification of GR. Usually, the gravitational lensing can be divided into weak and strong effects, depending on the amount of light deflection.  The weak deflection limit occurs when light passes far from the source, while the strong deflection limit occurs when  light passes close to the source. Gravitational lensing is a powerful astrophysical tool and profound research area in cosmology and gravitation, and various approaches have been proposed to investigate gravitational lensing in both limits.

Historically, the theory of gravitational lensing with the weak field approximation was developed first and it has been successfully employed to explain the physical observations \cite{schneider1992gravitational}. Various methods of weak gravitational lensing have been developed and applied to study BH physics. For instance, the deflection angle and weak gravitational lensing of the Reissner-Nordstr\"{o}m BH have been investigated in Ref. \cite{Sereno:2003nd}  {using} a perturbative method. The Taylor expansion formula was proposed for weak-field limit of static and spherically symmetric BH in \cite{Keeton:2005jd} and generalized to Kerr BH \cite{Sereno:2006ss,Werner:2007vu}. The Gibbons-Werner method was proposed in \cite{Gibbons:2008rj} where the Gauss-Bonnet theorem was applied. For more details on weak lensing, please see~\cite{schneider1992gravitational,Kayser:1992iyz,Bartelmann:1999yn,Weinberg:2013agg} for reviews.

In the last few decades, strong gravitational lensing has gained much attention. It has become a useful tool in detecting  {non-luminous} objects such as  {extrasolar} planets, dark matter, etc. Another motivation to study the strong gravitational lensing is the possibility of investigating the near horizon regions of BHs, by using the properties of the relativistic images. To our knowledge, Darwin \cite{darwin1959gravity} did the first work of strong lensing by compact objects with photon  {spheres,} like BHs and naked singularity. Virbhadra and Ellis \cite{Virbhadra:2002ju} investigated strong gravitational lensing and obtained the lens equation. Frittelli et al. \cite{Frittelli:1999yf} found an exact lens equation and integral expressions for its solutions. Bozza first proposed an analytical logarithmic expansion method for strong lensing \cite{Bozza:2002zj}. This method is universal and can be applied to any asymptotically flat spacetime.  {Afterwards}, various lensing observables for static and spherically spacetimes have been proposed \cite{Perlick:2003vg} and then widely studied for various spacetime such as Reissner-Nordstr\"{o}m BH by Eiroa et al. \cite{Eiroa:2002mk}, charged BHs with string theory \cite{Bhadra:2003zs}, brane-world BHs \cite{Whisker:2004gq}, rotating regular BHs \cite{Jusufi:2018jof}, phantom BHs \cite{Ovgun:2018prw}, Kerr and Kerr-Newman black holes \cite{Hsieh:2021scb,Hsieh:2021rru}, a  {holonomy-corrected} Schwarschild BH \cite{Junior:2024vdk} to mention a few. For more details, please see the review \cite{Bozza:2010xqn}. On the other hand, since modified theories of gravitation must agree with GR in the weak-field limit, in order to show deviations from GR it is necessary to probe strong fields. Hence, strong gravitational lensing is one of the most promising grounds where the theory of gravitation can be tested in its full form.

The paper is organized as follows. In Sec.~\ref{2s}, after briefly introducing the two quantum-modified metrics, we shall analyze the null geodesic equation in the equatorial plane of the BH. Then the deflection angle in the weak-field limit is calculated in Sect.~\ref{3s}. In Sec.~\ref{4s}, we focus on the light deflection in the  {strong-field} limit  {and discuss finite-distance corrections}, and the observables are presented in Sec.~\ref{5s}. Conclusions and discussions are made in the last section. We will use geometric units with $G=c=1$ unless we restore them to calculate observables.

\section{Theory}\label{2s}

\subsection{Black hole metrics}

In addressing the longstanding issue of general covariance in spherically symmetric gravity, the authors of Ref.~\cite{Zhang:2024khj} derived the equations for the effective Hamiltonian constraint, yielding two candidates dependent on a quantum parameter. The two candidates for effective Hamiltonian constraints generate two distinct quantum-modified metrics, each with a unique spacetime structure. The quantum-modified BH models are described by the following metrics \cite{Zhang:2024khj}
\begin{eqnarray}
    ds^2_{1}&=&-A_1(r)dt^2+B_1(r)dr^2+r^2d\Omega^2,\nonumber\\
    ds^2_{2}&=&-A_2(r)dt^2+B_2(r)dr^2+r^2d\Omega^2,\nonumber\\
    A_1(r)&=&1-\frac{2M}{r}+\zeta_M^2\frac{M^2}{r^2}\left(1-\frac{2M}{r}\right)^2,\nonumber\\
    A_2(r)&=&1-\frac{2M}{r},\nonumber\\
    B_1(r)&=&B_2(r)=A_1(r)^{-1},
    \label{metric0}
\end{eqnarray}
where $ds^2_{1}$ and $ds^2_{2}$ are the first and second quantum-modified metrics corresponding to the two effective Hamiltonian constraints, respectively. 
 {In this paper, we refer to $ds^2_{1}$ and $ds^2_{2}$  simply as model 1 and model 2, and use the subscripts 1 and 2 to denote quantities related to models 1 and 2, respectively.} Here $M$ is the ADM mass and $\zeta_M=\sqrt{4\sqrt{3}\pi\gamma^3l_p^2}/M$ is the rescaled quantum parameter, where $\gamma$ is the Barbero–Immirzi parameter and  $l_p$
is the Planck length.  {Since the quantum parameter is rescaled by the BH mass, we set the range of $\zeta_M$ from 0 to unity in the present work.} It can be readily verified that the above metrics exhibit the same relationship between mass $M$ and horizon radius $r_h$ as the classical BH, namely $r_h=2M$. The quantum-modified BH models also exhibit the same photon sphere radius as Schwarzschild BH, namely $r_{ps} = 3M$. This facilitates a direct comparison of the results obtained for the quantum-modified BH models with those of the classical BH.

\subsection{Setting the stage}
\begin{figure}[t]
    \centering
    \includegraphics[width=0.6\textwidth]{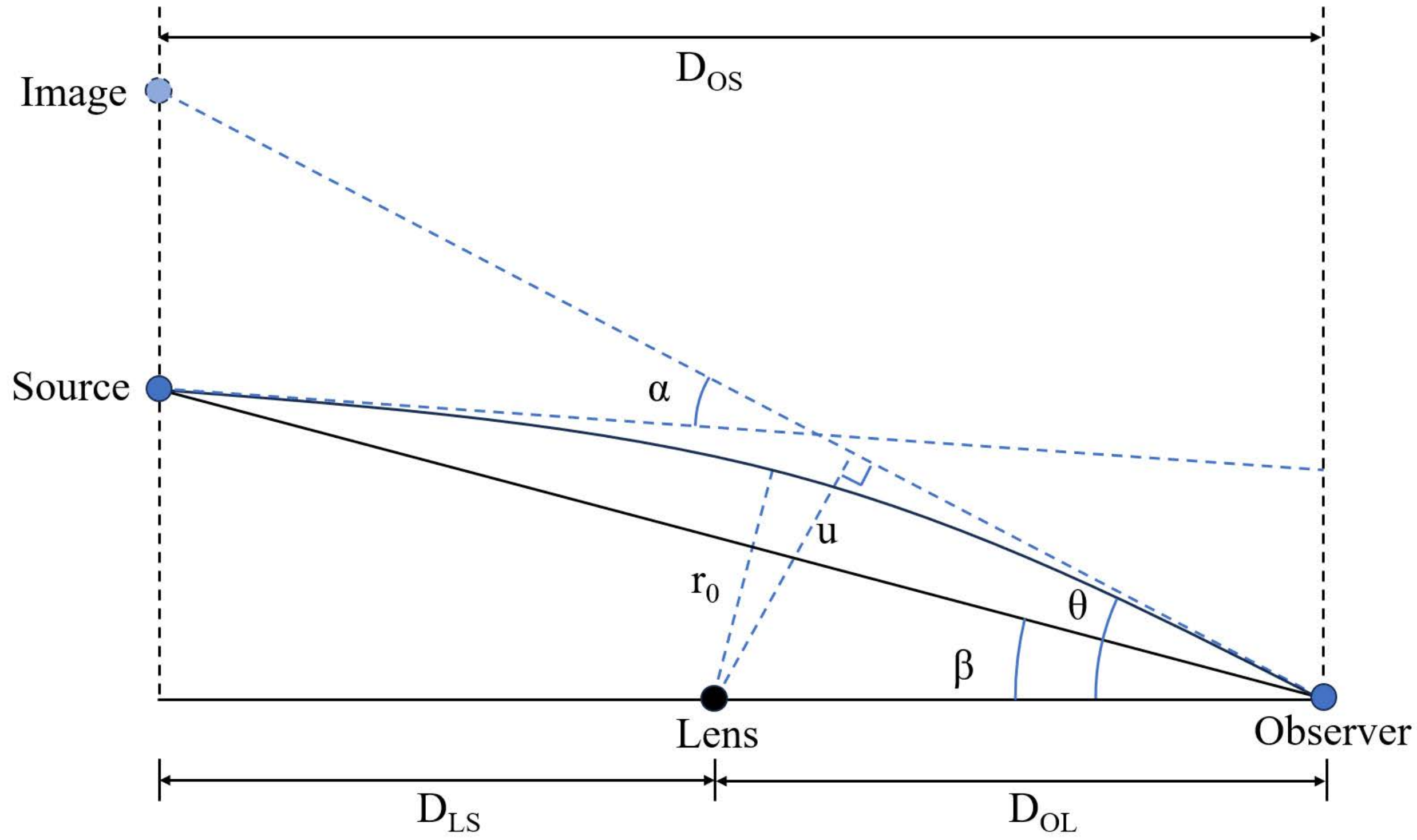}
    \caption{\it Schematic diagram of the lensing geometry.}
    \label{lensFig}
\end{figure}

To facilitate the comparison of deflection angles and other observables between the classical and quantum-corrected BHs, we first assume a static, spherically symmetric spacetime described as 
\begin{eqnarray}
    ds^2&=&-A(r)dt^2+B(r)dr^2+C(r)d\Omega^2.
    \label{metric1}
\end{eqnarray}
A schematic diagram of the lensing geometry is shown in Fig.~\ref{lensFig}, where $\beta$ is the angular position of the source, $\theta$ is the angular position of the image, $D_{OL}$ is the distance between the observer and the lens, and $D_{LS}$ is the distance between the lens and the source. The light ray originates from the source in the asymptotically flat region and is deflected by the BH, which acts as a lens, before reaching the observer in the flat region. Then the deflection angle $\alpha$ is given by \cite{weinberg1972,Virbhadra:1998dy,Cai:2023ite}
\begin{eqnarray}
    \alpha(r_0) = 2\int^\infty_{r_0} \frac{1}{C}\sqrt{\frac{AB}{1/u^2-A/C}}dr-\pi,
    \label{DA}
\end{eqnarray}
where $r_0$ is the closest distance to the lens during the propagation of the light and $u$ is the impact parameter. The impact parameter is related to $r_0$ through the following equation
\begin{eqnarray}
    u\equiv  \sqrt{\frac{C(r_0)}{A(r_0)}}.
    \label{ur0}
\end{eqnarray}
Generally, the integral \eqref{DA} does not admit an analytical solution, except in a few simple cases. Fortunately, several efficient methods (numerical or analytical) have been proposed for calculating the deflection angle in the weak-field limit ($r_0\gg M$)  or in the strong-field limit ($r_0\sim M$). The following sections will employ an approximate method \cite{Keeton:2005jd,Junior:2023xgl} to compute the deflection angle in the weak-field limit and Bozza's approach \cite{Bozza:2002zj} to determine the deflection angle in the strong-field limit.

\section{Deflection angles in the weak-field limit}\label{3s}
In this section, we shall compare the deflection angle between the classical and the two quantum-modified BH models in the weak-field limit ($r_0\gg M$). 

First, let us calculate the deflection angle $\alpha_1$ of the quantum-modified model 1. Substituting the metric \eqref{metric0} into the integral \eqref{DA}, we have
\begin{eqnarray}
    \alpha_1(r_0) &= &2\int^1_{0} \frac{dx}{\sqrt{(1-2h-x^2+2hx^3)[1+h^2(1+x^2)\zeta_M^2-2h^3(1+x^3)\zeta_M^2]}}-\pi,\nonumber\\
    &&
    \label{da10}
\end{eqnarray}
where we made the transformations $x=r_0/r$ and $h=M/r_0$ and used the relation \eqref{ur0}. Doing the Taylor expansion of the integral \eqref{da10} with respect to $h\ll 1$ and integrating it order by order, we obtain the following series
\begin{eqnarray}
    \alpha_1(r_0) &=&4h+\left(-4+\frac{15\pi}{4}-\frac{3\pi\zeta_M^2}{4}\right)h^2+\left(\frac{122}{3}-\frac{15\pi}{2}-\frac{10\zeta_M^2}{3}+\frac{3\pi\zeta_M^2}{2}\right)h^3\nonumber\\
    &&+\left(-130+\frac{3465\pi}{64}+10\zeta_M^2-\frac{105\pi\zeta_M^2}{32}+\frac{57\pi\zeta_M^4}{64}\right)h^4 +\mathcal{O}(h)^5.
    \label{da11}
\end{eqnarray}
Similarly, the deflection angle of $ds_2^2$ is
\begin{eqnarray}
    \alpha_2(r_0) &=&4h+\left(-4+\frac{15\pi}{4}-\frac{\pi\zeta_M^2}{4}\right)h^2+\left(\frac{122}{3}-\frac{15\pi}{2}-\frac{4\zeta_M^2}{3}+\frac{\pi\zeta_M^2}{2}\right)h^3\nonumber\\
    &&+\left(-130+\frac{3465\pi}{64}+4\zeta_M^2-\frac{45\pi\zeta_M^2}{32}+\frac{9\pi\zeta_M^4}{64}\right)h^4 +\mathcal{O}(h)^5.
    \label{da21}    
\end{eqnarray}
In the above equations, we retain only up to the fourth order. While this approach can be readily extended to higher orders, the fourth order is sufficiently accurate for our current purposes. It can be easily verified that Eqs.~\eqref{da11} and \eqref{da21} reduce to the case of Schwarzschild BH \cite{Junior:2023xgl} when the quantum parameter vanishes $\zeta_M=0$.
 
We wish to obtain the deflection angle as a function of the impact parameter $u$. To this end, our next step is to determine the dependence of $r_0$ on $u$. Inserting Eq.~\eqref{metric0} into Eq.~\eqref{ur0} gives the impact parameter of $ds^2_1$
\begin{eqnarray}
    \frac{u}{r_0} = \frac{1}{\sqrt{1-\frac{2M}{r_0}+\zeta_M^2\frac{M^2}{r^2_0}\left(1-\frac{2M}{r_0}\right)^2}}.
\end{eqnarray}
From the above equation, we can find $r_0$ as a series of $M/u$ for $ds_1^2$
\begin{eqnarray}
    \frac{r_0}{u} &=& 1-\frac{M}{u}-\left(\frac{3}{2}-\frac{\zeta_M^2}{2}\right)\left(\frac{M}{u}\right)^2-4\left(\frac{M}{u}\right)^3 \nonumber\\
    &&-\left(\frac{105}{8}-\frac{3\zeta_M^2}{4}+\frac{5\zeta_M^4}{8}\right)\left(\frac{M}{u}\right)^4+\mathcal{O}\left(\frac{M}{u}\right)^5.
    \label{r0u}
\end{eqnarray}
Thus, the deflection angle $\alpha_1(u)$ of the quantum-modified model 1 in the weak-field limit can be obtained by substituting Eq.~\eqref{r0u} into Eq.~\eqref{da11},
\begin{eqnarray}
    \alpha_1(u)& =&4\left(\frac{M}{u}\right)+\left(\frac{15\pi}{4}-\frac{3\pi\zeta^2_M}{4}\right)\left(\frac{M}{u}\right)^2+\left(\frac{128}{3}-\frac{16\zeta^2_M}{3}\right)\left(\frac{M}{u}\right)^3\nonumber\\
    &&+\left(\frac{3465\pi}{64}-\frac{225\pi\zeta_M^2}{32}+\frac{105\pi\zeta_M^4}{64}\right)\left(\frac{M}{u}\right)^4+\mathcal{O}\left(\frac{M}{u}\right)^5.
    \label{da1}
\end{eqnarray}
Similarly, we can calculate the deflection angle $\alpha_2(u)$ of model 2 in the weak-field limit. The result is
\begin{eqnarray}
    \alpha_2(u)& =&4\left(\frac{M}{u}\right)+\left(\frac{15\pi}{4}-\frac{\pi}{4}\zeta^2_M\right)\left(\frac{M}{u}\right)^2+\left(\frac{128}{3}-\frac{4}{3}\zeta^2_M\right)\left(\frac{M}{u}\right)^3\nonumber\\
    &&+\left(\frac{3465\pi}{64}-\frac{45\pi\zeta_M^2}{32}+\frac{9\pi\zeta_M^4}{64}\right)\left(\frac{M}{u}\right)^4+\mathcal{O}\left(\frac{M}{u}\right)^5.
    \label{da2}
\end{eqnarray}

\begin{figure}[t]
    \centering
    \includegraphics[width=0.6\textwidth]{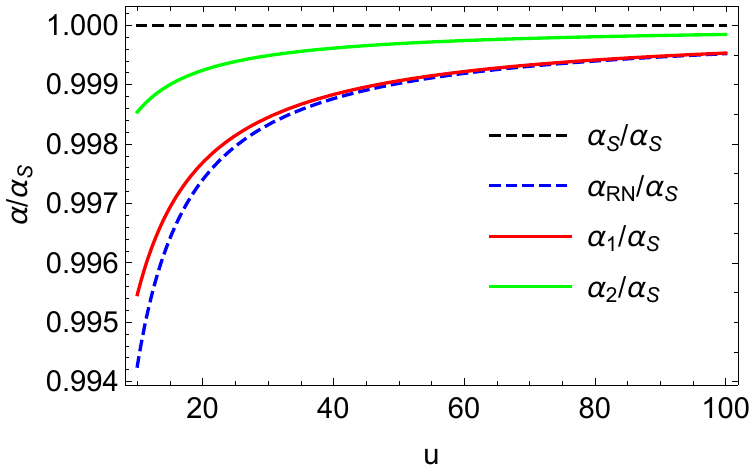}
    \caption{\it The deflection angles $\alpha_1$ and $\alpha_2$ in the weak-field limit as functions of the impact parameter $u$. $\alpha_S$ is the deflection angle of Schwarzschild BH. $\alpha_{RN}$ denotes the Reissner-Nordstr\"{o}m BH with $q=0.4$. $\alpha_1$ and $\alpha_2$ are the deflection angles of models 1 and 2 with $\zeta_M=0.4$, respectively. Here $M=1$ and all results are divided by $\alpha_S$ for better presentation.}
    \label{DA0WFL}
\end{figure}

\begin{figure}[t]
    \centering
    \includegraphics[width=0.6\textwidth]{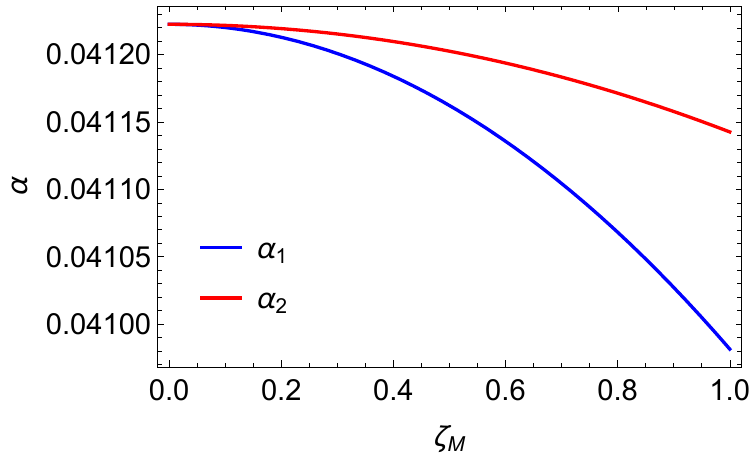}
    \caption{\it The deflection angles $\alpha_1$ and $\alpha_2$ in the weak-field limit as functions of the quantum parameter $\zeta_M$. Here we fix $M=1$ and $u=100$.}
    \label{DA12WFL}
\end{figure}

These results reduce to the case of Schwarzschild BH when $\zeta_M\rightarrow 0$. Notice that the first two terms of $\alpha_1(u)$ correspond precisely to the result for the Reissner-Nordstr\"{o}m BH \cite{Sereno:2003nd}, as long as we make the substitution $\zeta_M\rightarrow q$, where $q$ denotes the electric charge of the BH. This suggests that the quantum parameter $\zeta_M$ plays a role analogous to the electric charge in weak gravitational lensing. This similarity is easily explained by the fact that the quantum-modified metric $ds_1^2$ and the Reissner-Nordstr\"{o}m BH metric are equivalent up to the second-order post-post-Newtonianr expansion \cite{Keeton:2005jd}. Also notice that the higher-order coefficients in Eqs.~\eqref{da1} and \eqref{da2}  {become} negative when $\zeta_M$ is greater than some certain values. Thus, the deflection angles could be negative for large $M/u$. This should not bother us, because Eqs.~\eqref{da1} and \eqref{da2} are derived in the weak-field limit, which also implies $M/u\ll 1$. Nevertheless, the  {formulae} \eqref{da1} and \eqref{da2} for the deflection angle in the weak-field limit  {remain} a valuable tool for  {distinguishing} the classical BHs and the two quantum-modified BH models with large $u/M$. In Fig.~\ref{DA0WFL}, we plot a typical comparative analysis of the deflection angles exhibited by different BHs.  {In addition, the deflection angles in the weak-field limit decrease as $\zeta_M$ increases, and the deflection angle of $ds_2^2$ is always larger than that of $ds_1^2$ (see Fig.~\ref{DA12WFL}).}

\section{Deflection angles in the strong-field limit and finite-distance corrections}\label{4s}
In this section, we will discuss the deflection angles of the quantum-modified BH models in the strong-field limit ($r_0\sim M$), employing the scheme proposed by Bozza \cite{Bozza:2002zj}. The main idea of Bozza's scheme is to perform an expansion of the deflection angle near the radius of the photon sphere $r_{ps}$ and then give an analytical expression. The radius of the photon sphere $r_{ps}$ is calculated by \cite{Claudel:2000yi,Virbhadra:2002ju}
\begin{eqnarray}
    \frac{C'(r)}{C(r)} = \frac{A'(r)}{A(r)}.
    \label{rps}
\end{eqnarray}

Let us first rewrite the deflection angle \eqref{DA} as 
\begin{eqnarray}
    \alpha(r_0) &=& I(r_0) -\pi \nonumber\\
    I(r_0) &=&  2\int^\infty_{r_0} \frac{1}{C}\sqrt{\frac{AB}{1/u^2-A/C}}dr .
\end{eqnarray}
By introducing an auxiliary variable $z\equiv 1-\frac{r_0}{r}$, the integral $I(r_0)$ becomes
\begin{eqnarray}
    I(r_0) &=& \int^1_0 R(z,r_0) h(z,r_0) dz,\nonumber\\
     R(z,r_0)&=&\frac{2r^2\sqrt{A(r)B(r)C_0}}{r_0C(r)},\nonumber\\
    h(z,r_0)&=&\frac{1}{\sqrt{A_0-\frac{A(r)C_0}{C(r)}}}.
    \label{I0}
\end{eqnarray}
where $A_0\equiv A(r_0)$ and $C_0\equiv C(r_0)$. Notice that $R(z,r_0)$ is regular for any values of $z$ and $r_0$ and $h(z,r_0)$ diverges as $z\rightarrow 0$. To handle the divergent term $h(z,r_0)$, we expand the term under the square root and obtain
\begin{eqnarray}
    h(z,r_0) \sim h_0(z,r_0) = \frac{1}{\sqrt{m(r_0)z+n(r_0)z^2}},
    \label{h0}
\end{eqnarray}
where
\begin{eqnarray}
    m(r_0) &=& -r_0A'_0 +\frac{r_0A_0C'_0}{C_0},\nonumber\\
    n(r_0) &=& r_0\frac{-2r_0A_0{C'_0}^2-C_0^2(2A'_0+r_0A''_0)+C_0[2r_0A'_0C'_0+A_0(2C'_0+r_02C''_0)]}{2C_0^2}.\nonumber\\
\end{eqnarray}
Using Eq.~\eqref{h0}, we separate the divergent part from the integral \eqref{I0} as
\begin{eqnarray}
    I(r_0)&=&I_D(r_0)+I_R(r_0),\nonumber\\
    I_D(r_0)&=& \int^1_0 R(0,r_{ps}) h_0(z,r_0) dz,\nonumber\\
    I_R(r_0)&=& \int^1_0 [R(z,r_0) h(z,r_0)-R(0,r_{ps}) h_0(z,r_0)]dz,\nonumber\\
\end{eqnarray}
where $I_D$ and $I_R$ correspond to the divergent and regular parts of the integral, respectively. The divergent part $I_D$ can be integrated analytically and the result is
\begin{eqnarray}
    I_D(r_0) = \frac{2R(0,r_{ps})}{\sqrt{n(r_0)}} \log\left(\frac{\sqrt{n(r_0)}+\sqrt{m(r_0)+n(r_0)}}{\sqrt{m(r_0)}}\right).
    \label{ID0}
\end{eqnarray}
When $r_0\rightarrow r_{ps}$, $m(r_0)\rightarrow 0$ and thus $I_D(r_0) \rightarrow \infty$. By expanding Eq.~\eqref{ID0} near the photon sphere, we obtain
\begin{eqnarray}
    I_D(r_0) = - \frac{R(0,r_{ps})}{\sqrt{n(r_{ps})}}\log\left[\frac{1}{2}\left(\frac{r_0}{r_{ps}}\right)-1\right]+\mathcal{O}(r_0-r_{ps}).
\end{eqnarray}
Similarly, we expand the regular part $I_R$ near the photon sphere and obtain
\begin{eqnarray}
    I_R(r_0) = I_R(r_{ps}) +\mathcal{O}(r_0-r_{ps}).
\end{eqnarray}
Finally, the deflection angle $\alpha(r_0)$ becomes
\begin{eqnarray}
    \alpha(r_0) = - \frac{R(0,r_{ps})}{\sqrt{n(r_{ps})}}\left[\log\left(\frac{r_0}{r_{ps}}-1\right)-\log2\right]+I_R(r_{ps})-\pi+\mathcal{O}(r_0-r_{ps}),
    \label{dar0}
\end{eqnarray}
where the integral $I_R(r_{ps})$ can be computed numerically. It is convenient to rewrite the deflection angle as a function of the impact parameter $u$. Also expand the impact parameter $u$ near the photon sphere
\begin{eqnarray}
    u &=& u_{ps}+\Tilde{c}(r_0-r_{ps})^2, \nonumber\\
    \Tilde{c} &=& \frac{A(r_{ps})C''(r_{ps})-C(r_{ps})A''(r_{ps})}{4\sqrt{A(r_{ps})^3C(r_{ps})}},\nonumber\\
    u_{ps} &=& \frac{C(r_{ps})}{A(r_{ps})}.
\end{eqnarray}
Employing the above relation, Eq.~\eqref{dar0} can be reexpressed as
\begin{eqnarray}
    \alpha(u) &=& - \frac{R(0,r_{ps})}{2\sqrt{n(r_{ps})}}\left[\log\left(\frac{u}{u_{ps}}-1\right)+\log\left(\frac{u_{ps}}{\Tilde{c}r_{ps}^2}\right)-\log4\right] \nonumber\\
    &&+I_R(r_{ps})-\pi+\mathcal{O}(r_0-r_{ps}), \nonumber\\
    &=&-\bar{a}\log_{}{\left ( \frac{u}{u_{ps}}-1 \right )  }+\bar{b}+\mathcal{O}(r_0-r_{ps}),
    \label{dau}
\end{eqnarray}
where
\begin{eqnarray}
    \bar{a}&=&\frac{R(0,r_{ps})}{2\sqrt{n(r_{ps})}},\nonumber\\
    \bar{b}&=& - \bar{a}\left[\log\left(\frac{u_{ps}}{\Tilde{c}r_{ps}^2}\right)-\log4\right] +I_R(r_{ps})-\pi.
\end{eqnarray}
Using the relation $u=\theta D_{OL}$, the formula of the deflection angle can be rewritten as a function of $\theta$
\begin{eqnarray}
    \alpha(\theta)=-\bar{a}\log_{}{\left ( \frac{\theta D_{OL}}{u_{ps}}-1 \right )  }+\bar{b}.
    \label{datheta}
\end{eqnarray}
Here we neglect the error term for simplicity, since it does not affect our results. The specifics of this error term can be found in Refs.~\cite{Iyer:2006cn} and \cite{Tsukamoto:2016qro}.

\begin{figure*}[t!]
\centering
\includegraphics[width=0.47\textwidth]{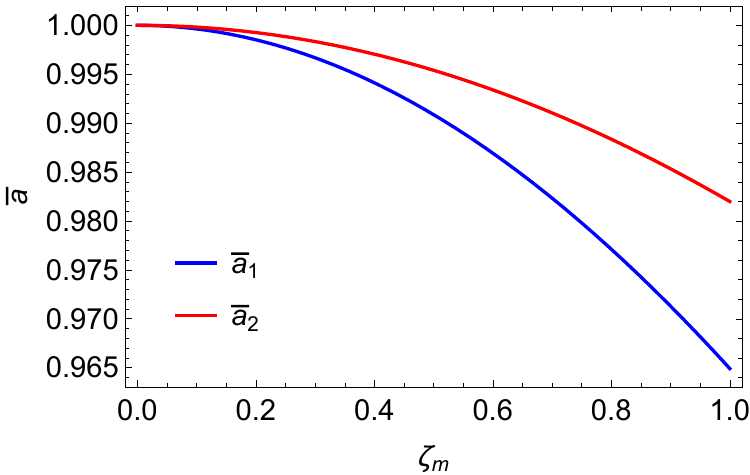}
\hfill%
\includegraphics[width=0.47\textwidth]{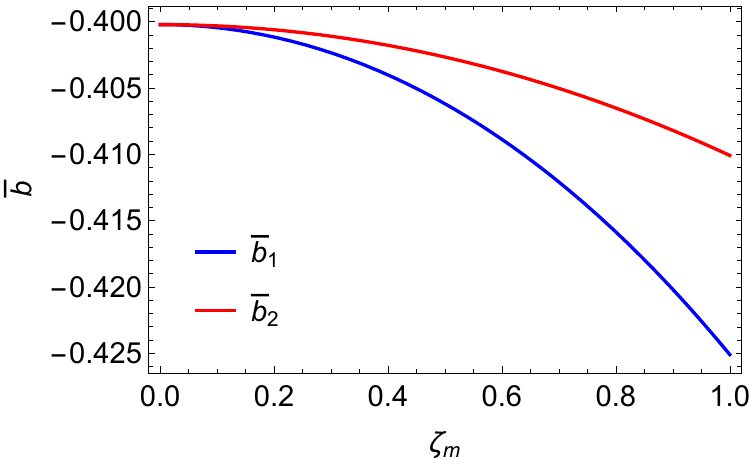}
\caption{\it The strong field coefficients  $\bar{a}$ (\textit{Left}) and $\bar{b}$ (\textit{Right}) of models 1 and 2 as functions of the quantum parameter.}\label{abarbbar}
\end{figure*}

Substituting the quantum-modified metrics \eqref{metric0} into the formula of the deflection angle \eqref{dau}, we can obtain the deflection angles of the quantum-modified models as functions of the impact parameter $u$ and the quantum parameter $\zeta_M$. The formulae for $u_{ps}$ of different quantum-modified models are
\begin{eqnarray}
    u_{ps1 }&=& \frac{27M}{\sqrt{27+\zeta_M^2}}, \quad\quad {\rm for} \quad ds_1^2,\nonumber\\
    u_{ps2} &=& 3\sqrt{3}M,\quad\quad \quad \quad{\rm for}\quad  ds_2^2. 
    \label{ups}
\end{eqnarray}
 {Notice that $u_{ps1}$ depends on $\zeta_M$, whereas $u_{ps2}$ does not.}
For $ds_1^2$ we have
\begin{eqnarray}
    \bar{a}_1 &=& \frac{3\sqrt{3}}{\sqrt{27+2\zeta_M^2}},\nonumber\\
    \bar{b}_1 &=& - \bar{a}_1\left[\log\left(\frac{2}{3}\right)+\log\left(\frac{27+\zeta_M^2}{27+2\zeta_M^2}\right)-\log4\right] +I_R^{(1)}-\pi,
    \label{abbar1}
\end{eqnarray}
and for $ds_2^2$ we have
\begin{eqnarray}
    \bar{a}_2 &=& \frac{3\sqrt{3}}{\sqrt{27+\zeta_M^2}},\nonumber\\
    \bar{b}_2 &=& - \bar{a}_2\left[\log\left(\frac{2}{3}\right)-\log4\right] +I_R^{(2)}-\pi.
    \label{abbar2}
\end{eqnarray}
The strong field coefficients $\bar{a}_1$, $\bar{a}_2$, $\bar{b}_1$ and $\bar{b}_2$ as functions of $\zeta_M$ are plotted in Fig.~\ref{abarbbar}. As one can see, the strong field coefficients decrease with increasing quantum parameter, and the strong field coefficients for $ds_1^2$ are always greater than  {those} for $ds_2^2$.

\begin{figure}[t]
    \centering
    \includegraphics[width=0.6\textwidth]{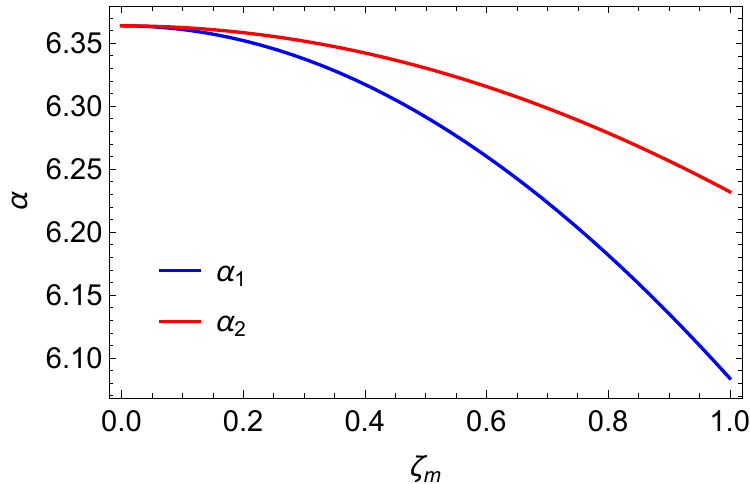}
    \caption{\it The deflection angles $\alpha_1$ and $\alpha_2$ in the strong-field limit as functions of the quantum parameter $\zeta_M$. Here we fix $M=1/2$ and $u=u_{ps}+0.003$.}
    \label{DA12SFL}
\end{figure}

When $M=1/2$ and $\zeta_M=0$, we have
\begin{eqnarray}
    \alpha_1(u) = \alpha_2(u) = -\log\left(\frac{2u}{3\sqrt{3}}-1\right)- 0.40023,
\end{eqnarray}
which recovers the formula of the deflection angle for Schwarzschild BH \cite{Bozza:2002zj}. As examples, the results for  $M=1/2$ and $\zeta_M=0.4$ are
\begin{eqnarray}
    \alpha_1(u) &=&  -0.99413\log\left(0.38604u-1\right)- 0.40403,\nonumber\\
    \alpha_2(u) &=& -0.99705\log\left(0.38490u-1\right)- 0.40179,
\end{eqnarray}
where $\alpha_1$ and $\alpha_2$ correspond to the deflection angles of the quantum-modified models $ds_1^2$ and $ds_2^2$, respectively. In Fig.~\ref{DA12SFL}, we plot the deflection angles  $\alpha_1$ and $\alpha_2$ in the strong-field limit as functions of the quantum parameter $\zeta_M$ for $M=1/2$ and $u=u_{ps}+0.003$. As we can see, the deflection angles for both quantum-modified models decrease with increasing quantum parameter, and model 1 exhibits a smaller deflection angle than model 2.

 {It is worth pointing out that we assumed the distances from the lens to the source and the observer are infinite in deriving the formula of the deflection angle in the strong-field limit. In the cases of Sgr A* and M87*, however, the source could be in the region near the supermassive BHs. Consequently, a finite-distance correction for the deflection angle must be taken into consideration \cite{Ishihara:2016vdc,Ishihara:2016sfv}. Following the formulation proposed by Ishihara \textit{et al.} \cite{Ishihara:2016sfv}, we consider the case where the source and the observer are far from the lens ($r_S\gg u$, $r_R\gg u$), while the light ray passes near the photon sphere. Here, $r_S$ and $r_R$ represent the distances from the lens to the source and the observer (receiver), respectively. As the result, we obtain the finite-distance corrections for the two quantum-modified BH models as
\begin{eqnarray}
    \delta\alpha_1 &=& -\frac{M}{u}\delta_R^2 -\frac{M}{u}\delta_S^2 - \frac{M^2}{2u^2}(1-\zeta_M^2)\delta_R^3- \frac{M^2}{2u^2}(1-\zeta_M^2)\delta_S^3+\mathcal{O}(\delta_R^4+\delta_S^4),\nonumber\\
    \delta\alpha_2 &=& -\frac{M}{u}\delta_R^2 -\frac{M}{u}\delta_S^2 - \frac{M^2}{2u^2}(1+\frac{\zeta_M^2}{3})\delta_R^3- \frac{M^2}{2u^2}(1+\frac{\zeta_M^2}{3})\delta_S^3+\mathcal{O}(\delta_R^4+\delta_S^4),
    \label{fdCorrections}
\end{eqnarray}
where $\delta_R=u/r_R$ and $\delta_S=u/r_S$. The above formulae suggest that the finite-distance corrections are of the order of $\mathcal{O}(\delta_R^2+\delta_S^2)$, and the influence of quantum parameters $\zeta_M$ on these corrections is at least of third-order. To estimate magnitude of these corrections, we use the data of mass and distance for SgrA* and M87* as examples: $M=4.4\times 10^6M_\odot$ and $r_R=D_{OL}=8.5$ Kpc for SgrA*, and $M=6.5\times 10^9M_\odot$ and $r_R=D_{OL}=16.8$ Mpc for M87*. Note that $r_R\sim10^{10}M$ for both SgrA* and M87*. Substituting $u=6M$, $r_S=10^6M$ (which corresponds to $\sim10^{-1}$ pc for SgrA* and $\sim10^{2}$ pc for M87*), and the above data into Eq.~\eqref{fdCorrections}, one can readily find that the finite-distance corrections are $\sim10^{-6}$ plus $\sim10^{-13}\zeta_M^2$ arcseconds for both SgrA* and M87* in both BH models. It is clear that the quantum parameter's contribution to the finite-distance corrections is quite small.
}

\section{Obervables in strong gravitational lensing by supermassive BHs}\label{5s}

Let us now investigate how the quantum parameter affects observables in strong gravitational lensing. Assuming that the source, lens and observer are almost aligned, the lens equation in the strong-field limit reads \cite{Bozza:2001xd} 
\begin{eqnarray}
    \beta=\theta-\frac{D_{LS}}{D_{OS}} \Delta \alpha_n,
    \label{lensEq}
\end{eqnarray}
where $\beta$ is the angular position of the source, $\theta$ is the angular position of the image, $D_{OL}$ is the distance between the observer and lens, $D_{LS}$ is the distance between the lens and source and $D_{OS}=D_{OL}+D_{LS}$. In strong gravitational lensing, light rays orbit around the lens $n$ times before reaching the observer, and thus we have $\Delta \alpha_n = \alpha-2n\pi$. Combining the deflection angle \eqref{datheta} and the lens equation \eqref{lensEq}, we can approximately obtain the expression for the $n$th angular position of the image \cite{Bozza:2002zj}
\begin{eqnarray}
    \theta _{n} \simeq \theta _{n}^{0}+\frac{u_{ps}e_{n}\left ( \beta-\theta _{n}^{0} \right )D_{OS} }{\bar{a}D_{LS}D_{OL}},
    \label{nposition}
\end{eqnarray}
where
\begin{eqnarray}
    \theta_{n}^{0}&=&\frac{u_{ps}}{D_{OL}}\left ( 1+e_{n} \right ), \nonumber\\ 
    e_{n}&=&e^{\frac{\bar{b}-2n\pi}{\bar{a}}}.
\end{eqnarray}
Here $\theta_{n}^{0}$ is the image position satisfying $\alpha(\theta_{n}^{0}) = 2n\pi$. Another important parameter is the magnification of $n$th image, and it can be evaluated as \cite{Bozza:2002zj}
\begin{eqnarray}
    \mu_{n}=\left ( \frac{\beta}{\theta }\frac{\partial \beta}{\partial \theta } {}   \right ) ^{-1} \bigg|_{\theta _{n}^{0}} =e_{n}\frac{u_{ps}\left ( 1+e_{n} \right )D_{OS} }{\bar{a}\beta D_{OL}^{2}D_{LS}}.
    \label{nmagnification}
\end{eqnarray}

\begin{figure*}[t!]
\centering
\includegraphics[width=0.47\textwidth]{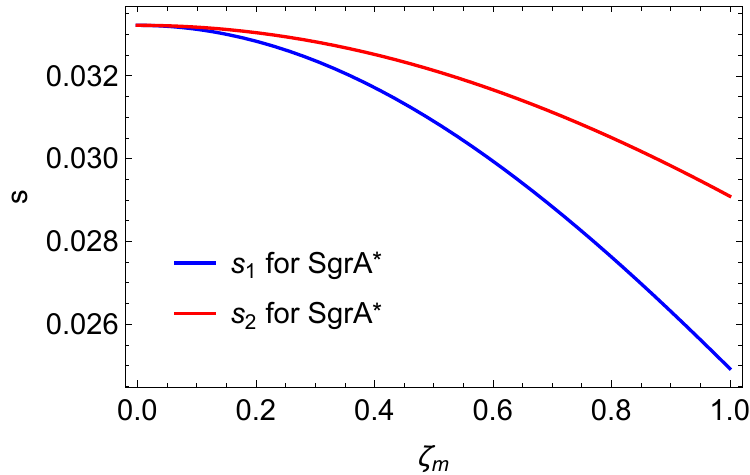}
\hfill%
\includegraphics[width=0.47\textwidth]{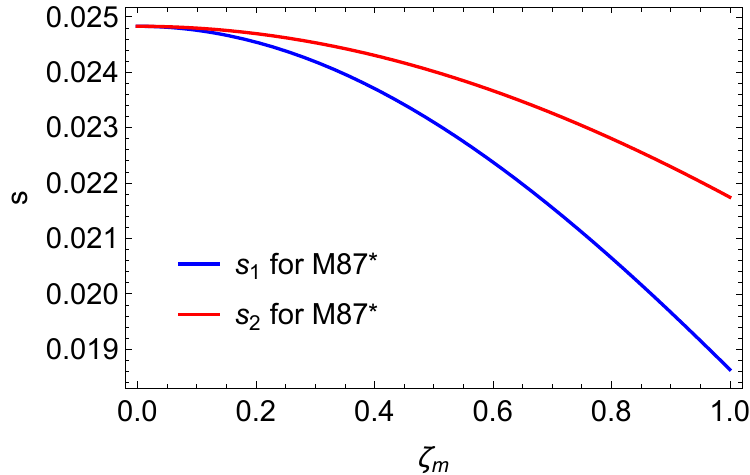}
\caption{\it Angular separations $s$ ($\mu$arcsecs) as functions of the quantum parameter for supermassive BHs SgrA* (\textit{Left}) and M87* (\textit{Right}) with different quantum-modified BH models.}\label{S12SFL}
\end{figure*}

Now the image position and the magnification relate to the strong field coefficients $\bar{a}$ and $\bar{b}$ through Eqs.~\eqref{nposition} and \eqref{nmagnification}. Next, let us solve the inverse problem, which involves calculating the strong field coefficients using the image position and the magnification. By doing so, we can deduce information about the lensing object from astronomical observations. Denote the  {outermost} image position as $\theta_1$, and pack all the remaining ones as $\theta_\infty$. Then the minimum impact parameter can be obtained as 
\begin{eqnarray}
    u_{ps} =D_{OL}\theta_\infty,
    \label{utheta}
\end{eqnarray}
where $\theta_\infty$ can be obtained from Eq.~\eqref{nposition} in the limit $n\rightarrow\infty$. Therefore, the observables can be defined as \cite{Bozza:2002zj,Bozza:2003cp}
\begin{eqnarray}
    s&=&\theta _{1}-\theta _{\infty }=\theta_{\infty}e^{\frac{\bar{b}-2\pi}{\bar{a}}},\nonumber\\
    \mathcal{R}&=&\frac{\mu_{1}}{\sum_{n=2}^{\infty}\mu_{n} }=e^{\frac{2\pi}{\bar{a}} },\nonumber\\
    \Delta T_{2,1}&=&2\pi\theta_{\infty}D_{OL},
    \label{observables}
\end{eqnarray}
where $s$ is the angular separation between the outermost and asymptotic relativistic images, $\mathcal{R}$ is the relative magnification of the outermost relativistic images and $\Delta T_{2,1}$ is the time delay between the first and second images. Inverting the above formulae gives
\begin{eqnarray}
    \bar{a} &=&  \frac{2\pi}{\log \mathcal{R}},\nonumber\\
    \bar{b} &=&  \bar{a}\log\left( \frac{s\mathcal{R}}{\theta_\infty} \right).
\end{eqnarray}

\begin{figure}[t]
    \centering
    \includegraphics[width=0.6\textwidth]{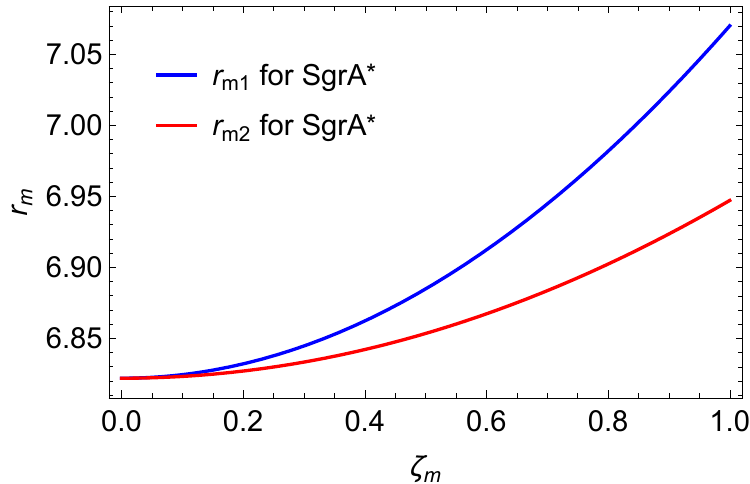}
    \caption{\it Relative magnifications as functions of the quantum parameter with different quantum-modified BH models. Here $r_{m}=2.5\log_{10}{\mathcal{R}}$.}
    \label{rm12SFL}
\end{figure}

\begin{figure*}[t!]
\centering
\includegraphics[width=0.47\textwidth]{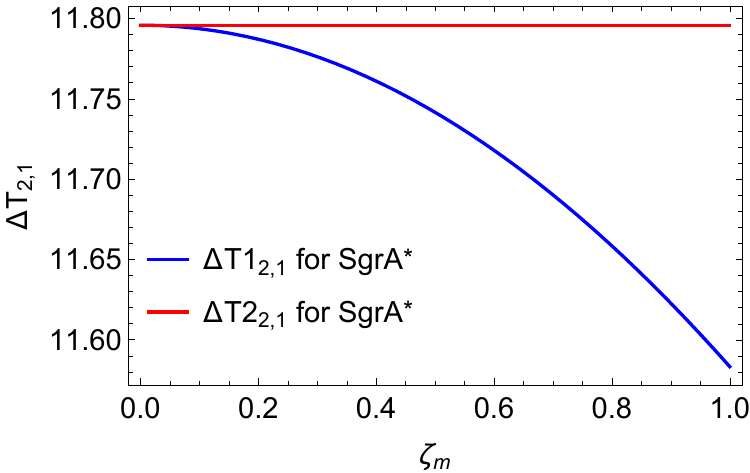}
\hfill%
\includegraphics[width=0.49\textwidth]{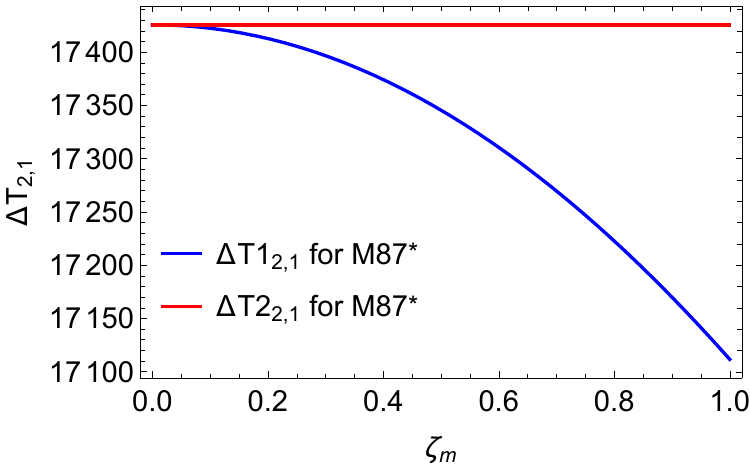}
\caption{\it Time delays $\Delta T_{2,1}$ (mins) as functions of the quantum parameter for supermassive BHs SgrA* (\textit{Left}) and M87* (\textit{Right}) with different quantum-modified BH models.}\label{t12SFL}
\end{figure*}

With these formulae \eqref{observables} in hand, let us now consider some realistic cases, such as SgrA* and M87*. Substituting the strong field coefficients (\eqref{abbar1} and \eqref{abbar2}) and  {the data for SgrA* and M87*} into the formulae \eqref{observables}, we can evaluate the observables for supermassive BHs SgrA* and M87* with different quantum-modified BH models. The results are presented in Figs.~\ref{S12SFL}-\ref{t12SFL}. It can be easily checked from these figures that all the results recover the Schwarzschild case when $\zeta_M=0$. In Fig.~\ref{S12SFL}, we show the angular separation as a function of the quantum parameter.  {As the quantum parameter increases, we observe that model 1 has a smaller angular separation compared to model 2.} The relative magnifications versus the quantum parameter are depicted in Fig.~\ref{rm12SFL}, which applies to both SgrA* and M87* because the relative magnification is independent of mass and distance.  {Figure~\ref{rm12SFL} reveals an increase in relative magnifications with the quantum parameter, accompanied by a growing difference between the two models.} As depicted in the Fig.~\ref{t12SFL}, the time delays  {for} model 2 remains constant and equal to the Schwarzschild case, while the time delays  {for} model 1 decreases with increasing quantum parameter. The behavior of the observables described above can be well understood by referring back to the behavior of the strong field coefficients in Eqs.~\eqref{abbar1}, \eqref{abbar2} and Fig.~\ref{abarbbar}.

\section{Conclusions and discussions}\label{6s}

In this work, we investigated the gravitational lensing effects of two quantum-modified BH models recently proposed in effective quantum gravity \cite{Zhang:2024khj}. The light deflection angles in both the weak-field limit and the strong-field limit were calculated via Keeton and Petters' approach \cite{Keeton:2005jd} and Bozza's approach \cite{Bozza:2002zj}, respectively. In the weak-field limit, the quantum correction decreases the deflection angles of the two models and increases their difference, with model 1's deflection angle always less than the model 2's.  {In the strong-field limit, the deflection angles of the two models decrease as $\zeta_M$ increases, with model 1's deflection angle always less than the model 2's.}  {We also discuss the finite-distance corrections for the deflection angle in both BH models by using the formulation proposed by Ishihara \textit{et al.} \cite{Ishihara:2016sfv}.}

Furthermore, we calculated the lensing observables in the strong-field limit, using the data of the supermassive BHs SgrA* and M87*.  {Notice that the behavior of the lensing observables depends on the minimum impact parameter and the strong field coefficients (see Eqs.~\eqref{utheta} and \eqref{observables}). Both $\bar{a}_1$ and $\bar{a}_2$ decrease as $\zeta_m$ increases, which explains the rapid growth of the relative magnification with increasing $\zeta_M$.} The slight differences between $\bar{a}_1$ and $\bar{a}_2$, as well as their independence of mass, are also reflected in the behavior of the relative magnification.  {For the angular separation, both models start from the Schwarzschild value and decrease at different rates with increasing $\zeta_m$, due to $\bar{b}_1$ and $\bar{b}_2$ being negative. As one can see from Eq.~\eqref{ups},} the minimum impact parameter for model 2 is constant, while for model 1, it decreases with $\zeta_M$. Therefore, the time delays  {for} model 2 remains constant and equal to the Schwarzschild case, while the time delays  {for} model 1 decreases with increasing $\zeta_m$. 

 {The present results suggest that the quantum parameter plays a role analogous to the electric charge in  weak gravitational lensing \cite{Sereno:2003nd,Keeton:2005jd}. In the strong-field limit, the effects of the quantum parameter on the deflection angle, the angular separation, and the relative magnification are opposite to those of the electric charge of Reissner-Nordstr\"{o}m BH \cite{Bozza:2002zj}, the scalar charge in Einstein-Maxwell-conformally coupled scalar theory \cite{QiQi:2023nex}, and the LQG parameters in Refs.~\cite{Alonso-Bardaji:2021yls,Alonso-Bardaji:2022ear,Soares:2023uup,Junior:2023xgl}.} In conclusion, our theoretical study shows the  {crucial difference} between the classical BHs and the two quantum-modified BH models, which may be useful in distinguishing between gravitational models in future astronomical observations.

 \vspace{1cm}

{\bf Acknowledgments}

We appreciate Mian Zhu and Wen-Cong Gan for helpful discussion. We especially thank Cong Zhang for helping us to clarify the description of LQG. This work is supported by the National Natural Science Foundation of China (NSFC) with Grant No.~12305064, No.~12365009 and No.~12205123 and Jiangxi Provincial Natural Science Foundation with Grant No. 20224BAB211020, No. 20232BAB201039 and No.~20232BAB211029.

 \vspace{1cm}

\clearpage

\bibliographystyle{apsrev4-1}
\bibliography{GR}

\end{document}